\documentclass[prepare]{elsarticle}
\usepackage{lineno,hyperref}
\usepackage{color}
\usepackage{textcomp}
\usepackage{gensymb}
\usepackage{CJKutf8}

\journal{Journal of \LaTeX\ Templates}

\begin{document}
\begin{CJK}{UTF8}{gkai}
\begin{frontmatter}

\title{\boldmath Design and performance testing of a T0 detector for the CSR External-target Experiment}


\author[address1,address2]{D.Hu\corref{correspondingauthor}}
\ead{hurongr@ustc.edu.cn}

\author[address1,address2]{X.Wang}

\author[address1,address2]{M.Shao\corref{correspondingauthor}}
\ead{swing@ustc.edu.cn}

\author[address1,address2]{Y.Zhou}
\author[address1,address2]{S.Ye}
\author[address1,address2]{L.Zhao}

\author[address1,address2]{Y.Sun}

\author[address1,address2]{J.Lu}

\author[address1,address2]{H.Xu}

\address[address1]{State Key Laboratory of Particle Detection and Electronics, University of Science and Technology of China, 96 Jinzhai Road, Hefei 230026, China}
\address[address2]{Department of Modern Physics, University of Science and Technology of China (USTC), 96 Jinzhai Road, Hefei 230026, China}
\cortext[correspondingauthor]{Corresponding author}


\begin{abstract}
The Cooling Storage Ring (CSR) External-target Experiment (CEE) at the Heavy Ion Research Facility in Lanzhou (HIRFL), China, is the first multi-purpose  nuclear physics experimental device to operate in the Giga electron-volt (GeV)  energy range.  The primary goals of the CEE are to study the bulk properties of dense matter and to understand the quantum chromo-dynamic (QCD) phase diagram by measuring the charged particles produced in heavy-ion collisions in the target region with a large acceptance. The CEE is a spectrometer that focuses on charged final-state particle measurements running on the HIRFL-CSR. The time-of-flight (TOF) system is critical for identifying charged particles in the GeV energy region. In the CEE spectrometer, the TOF system consists of three parts: T0, internal TOF, and external TOF, which are used for the final-state particle identification. The T0 detector provides a high-precision start time for the TOF system by measuring the crossing time of the heavy ion beam. This study details the design,  performance simulation, and performance testing of the T0 detector. The simulation results and heavy-ion beam test show that the T0 detector prototype has an excellent time resolution, which is better than 30 ps, and fulfills the requirements of the CEE.
\end{abstract}


\end{frontmatter}

\section{Introduction}
Studying the bulk properties of strong-interaction matter and understanding quantum chromodynamics (QCD) phase diagrams have been hot topics and frontiers in nuclear physics research. Relativistic heavy-ion collision experiments such as ALICE and STAR  have made major breakthroughs in these research fields in recent years. In particular, the latest results of the energy scan program of the relativistic heavy-ion collider (RHIC) show that the critical point of the QCD phase transition and other new physics related to the phase structure may occur at low temperatures and high baryon densities\cite{adam2021nonmonotonic}. New accelerators and particle detectors are planned or under construction to carry out heavy-ion collision experiments in this energy region (several to tens of GeV )\cite{friese2006cbm,toneev2007nica}. Chinese nuclear physicists have proposed to use the existing nuclear device Heavy-Ion Research Facility-Cooling Storage Ring (HIRFL-CSR)\cite{xia2002heavy} and the under-construction High-Intensity heavy-ion Accelerator Facility (HIAF)\cite{yang2013high}, and build a multi-purpose spectrometer called the CSR  External-target Experiment (CEE) to study nuclear matter at high density.\par

\begin{figure}[htbp]
\centering
\includegraphics[width=.8\textwidth]{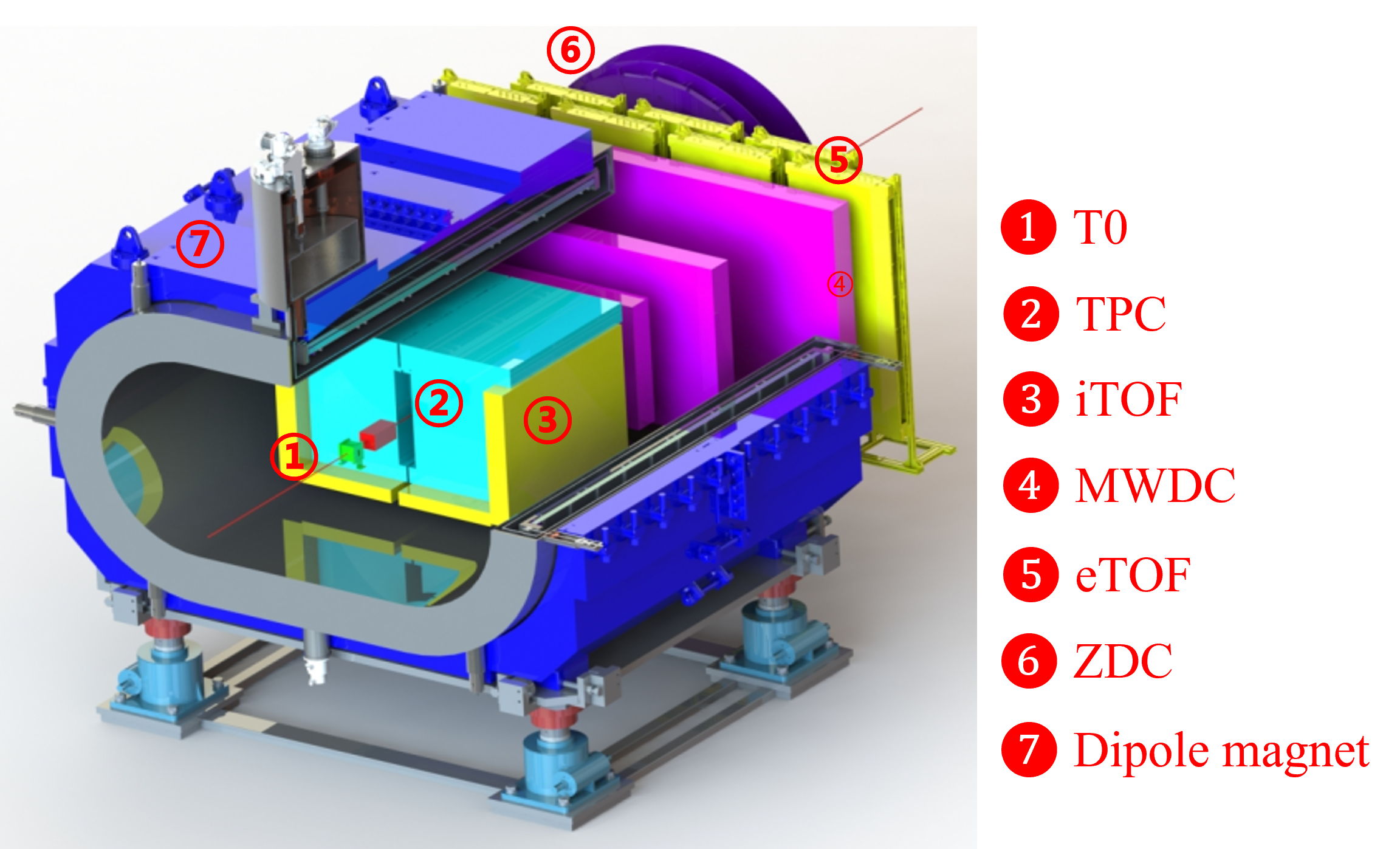}
\caption{ Schematic layout of the CEE spectrometer.}
\label{fig:1}
\end{figure}

The CEE is a spectrometer focusing on charged final-state particle measurement that runs on HIRFL-CSR and will continue to operate with possible upgrades at HIAF. It works in a fixed target mode and is designed to achieve high-precision measurement of charged particles with large detection acceptance.  Fig\ref{fig:1} shows the conceptual design diagram of the CEE. The particles identification (PID) of CEE relies on the TOF system, including the internal TOF (iTOF) in medium rapidity\cite{Wang_2022} (approximately 30-110$\degree$) and the external TOF (eTOF) in high rapidity\cite{Wang_2022} 
 (approximately 0-30$\degree$). A T0 detector is used to determine the start time for the TOF system by measuring the beam particle (projectile nucleus) arrival time. The T0 detector is placed in the beamline $\sim$ 30 cm upstream of the target, which is located inside the time projection chamber (TPC)\cite{li2016simulation}. The overall time resolution of the TOF system must be \textless 50 ps to achieve 3$\sigma$ $p/K/\pi$ identification under 2 GeV/c momentum for the TOF system, according to calculation and simulation. This means that if the intrinsic time resolution of the TOF detector is required to be \textless 40 ps, the T0 timing jitter should be no more than 30 ps\cite{Wang_2022}. 

\section{Module design}
\begin{figure}[htbp]
\centering
\includegraphics[width=.6\textwidth]{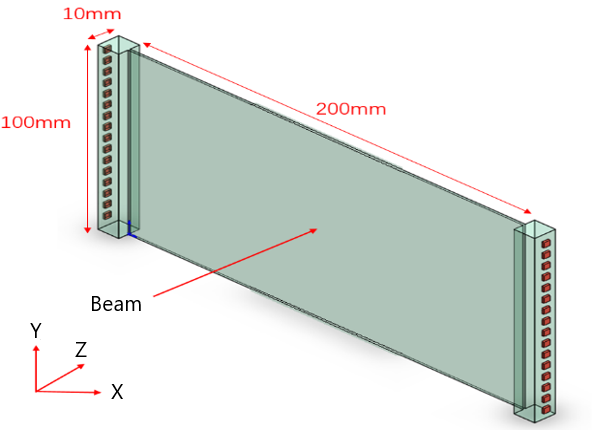}
\caption{Schematic diagram of the structure of the T0 detector.}
\label{fig:2}
\end{figure}

The T0 detector adopted a thin plastic scintillator coupled with the SiPMs readout. When the heavy-ion beam passed through the scintillator plate, energy was deposited through ionization, and fluorescence was emitted. The fluorescent photons reached the SiPMs after being transmitted through the scintillator and the light guide. Finally, time information was obtained by measuring the output signals of the SiPMs. Fig\ref{fig:2} shows a schematic of the T0 structure. The central part of the T0 was a thin plastic scintillator plate with an active area of 200 mm (X) $\times$ 100 mm (Y) and the thickness was 0.8–5 mm according to different types of beam nuclei. Both sides along the X (horizontal) direction of the scintillator were light guides with dimensions of 10 mm (X) $\times$ 100 mm (Y) $\times$ 10 mm (Z),  and a 5 mm groove was opened on each of the light guides to fix the central thin plastic scintillator plate. Optical glue was used to couple the scintillator plate and the light guide. On each side, 16 SiPMs were arranged evenly with a spacing of 6 mm. There was a 0.5 mm air gap between the SiPM window and the light guide. To achieve the required high-precision timing performance, no wrapping is applied to the outer surfaces of the scintillator, allowing only photons that satisfy the total internal reflection restriction to reach the SiPMs. In our design, the scintillator and light guide were both EJ-200 plastic scintillators because of their excellent properties, such as high light yield (10000/MeV), fast timing (decay time$\sim$2.1 ns), long light attenuation length ($\sim$380 cm), and good radiation tolerance. For the SiPMs, we chose Hamamatsu S13360-3025PE because of its high gain, excellent time resolution, and immunity to the effects of the magnetic field.\par

\section{Simulation study of T0 performance}
To study the performance of the T0 detector, we built a Monte Carlo simulation program based on GEANT4 ~\cite{agostinelli2003geant4} to evaluate the entire physical process, from energy deposition to final signal formation. In our simulations, the geometric parameters of the detector were set as described in the previous section, and the related parameters of the plastic scintillator and SiPM were set according to the official manual~\cite{Hamasipm}. The heavy-ion beam was incident from the centre of the scintillator along the Z direction (Fig~\ref{fig:2}), and the beam spot was set as a circle with a diameter of 5 mm. The influence of dispersion of the beam momentum ($\sim$0.1\%) was not considered in this simulation.

\subsection{Energy deposition and photon collection}
\begin{figure}[htbp]
\centering
\includegraphics[width=.6\textwidth]{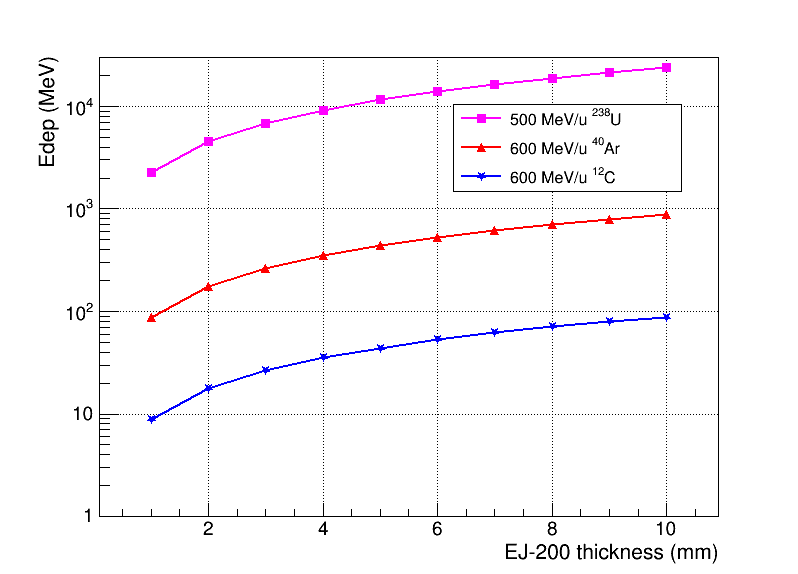}
\caption{ Total energy deposited by typical light, medium and heavy nucleon number beams in different thickness plastic scintillators in the HIRFL-CSR energy region.}
\label{fig:3}
\end{figure}

We studied the effect of total energy deposited in scintillators with different thicknesses for typical light (600 MeV/u $^{12}$C), medium (600 MeV/u $^{40}$Ar), and heavy (500 MeV/u $^{238}$U) nuclei in the HIRFL-CSR energy region. The simulation results are shown in Fig\ref{fig:3}. When these nuclei pass through a plastic scintillator of the same thickness,  $^{238}$U nuclei deposit much more energy than  $^{12}$C and $^{40}$Ar. For example, when the thickness of the plastic scintillator was 1 mm, the total energy deposited by $^{238}$U is 2270MeV, and the beam energy loss was 9.54 MeV/u.\par
The CEE experiment requires the beam energy loss in the T0 detector to be less than 10 MeV/u, so the material budget of T0 (including the plastic scintillator and light-tight wrapping) should be strictly controlled. Meanwhile, the scintillator should not be too thin to maintain mechanical strength and light collection efficiency. Taking all these considerations into account, a thickness of 0.8 mm was selected for the heavy nucleus beam (e.g., $^{238}$U).  The energy deposited in a 0.8 mm-thick scintillator by a medium-mass heavy-ion beam represented by $^{40}$Ar was $\sim$ 100 MeV, which can provide sufficient light output, as will be detailed in the following discussion. Therefore, a 0.8 mm-thick scintillator was also used for the medium nucleus beam. Finally, according to the simulation results in Fig~\ref{fig:3}， for light nuclear beams such as $^{12}$C， the energy deposition in the scintillator may reach 100 MeV only when the scintillator is as thick as 10 mm (in this case, the beam energy loss was \textless10 MeV/u). Since the quenching effect~\cite{birks1951scintillations} of $^{12}$C is smaller than that of $^{40}$Ar， the $^{12}$C beam produces more light outputs in the plastic scintillator than the $^{40}$Ar beam with the same energy deposition.  Therefore, for light nuclear beams, such as $^{12}$C, a scintillator with a thickness of 5 mm was chosen.\par

The energy deposited In the plastic scintillator excites scintillator molecules to emit photons. However, the relationship between the total number of emitted photons and the energy loss is not linear due to the quenching effect. An effective model to describe the quenching effect of heavy-ion beams in plastic scintillators is the BTV model, also known as the core-halo model~\cite{tarle1979cosmic,dwyer1985plastic}.  The model considers that the light yield can be divided into two regions, depending on the distance from the energy deposition location to the beam-crossing path. The inner region with a higher ionization energy loss density is called the core region, and the relationship between its light yield and energy loss can be described using the Birks formula ~\cite{birks1951scintillations}. The region further away from the core region is called the halo region, where the energy loss density is low, and its light yield is proportional to the energy loss (i.e., without quenching). The parameterization of the BTV model can be described using the following formula:
\begin{equation}
\frac{dL}{dx} = \frac{A(1-f_h)\frac{dE}{dx}}{1+B_s(1-f_h)\frac{dE}{dx}}+Af_h\frac{dE}{dx} 
\label{equation:BTV}
\end{equation}

where dL/dx is the light output per path length, $f_h$ is the ratio of energy deposited in the halo region to the total deposition energy, $B_s$ is the Birks constant, A is the global normalization coefficient, and dE/dx is the energy loss per unit of distance. Zhou Yong et al.~\cite{zhouyong2016dampe} measured the light output for various heavy-ion beams and summarised the fit parameters of Formula~\ref{equation:BTV} for the EJ-200 plastic scintillator. Based on these fit parameters and the simulation results in Fig. ~\ref{fig:3}, we calculated the quenching factor (the ratio of excited photon number with and without quenching effect) for 500 MeV/u $^{238}$U, 600 MeV/u $^{40}$Ar and 600 MeV/u $^{12}$C, which were 0.203, 0.284, and 0.615, respectively.\par

\begin{figure}[htbp]
\centering
\includegraphics[width=.6\textwidth]{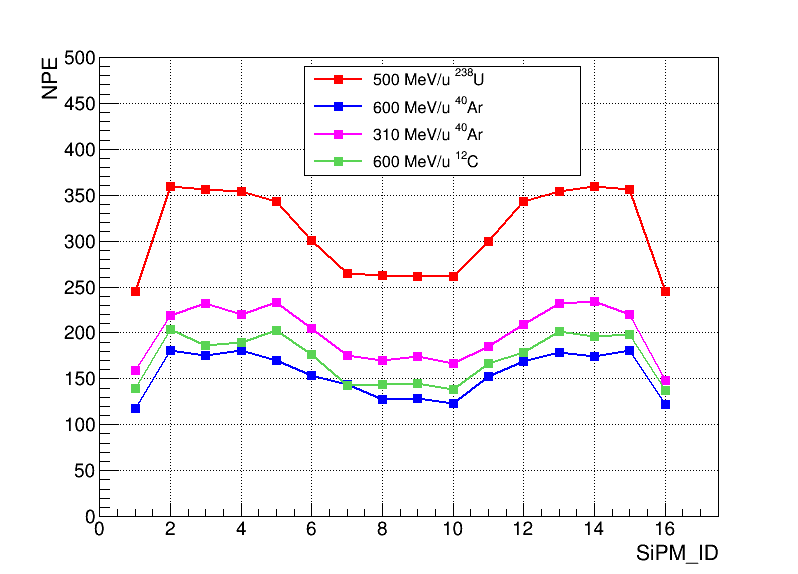}
\caption{The number of photoelectrons received by the 16 SiPM arrays on the side of T0}
\label{fig:4}
\end{figure}

The number of photoelectrons (NPE) received by the SiPM arrays can be obtained by substituting the calculated values into the simulation process. Table~\ref{fig:t1}  lists the calculation and simulation results for different nuclei beams. The last row shows the average number of photoelectrons received by a single SiPM. The simulation results of a lower-energy (310 MeV/u) $^{40}$Ar beam are also listed in Table~\ref{fig:t1}.Fig~\ref{fig:4} shows the number of photoelectrons received by each of the 16 SiPMs on one side of the T0 detector, where the result for 500 MeV/u $^{238}$U  was scaled down by a factor of 10. The horizontal ordinate in Fig.~\ref{fig:4} represents the SiPM ID at different positions, that is, SiPM 1, SiPM 2, ..., SiPM 16 along the Y-axis in Fig.~\ref{fig:2} It can be seen that the number of photoelectrons received by SiPMs at different positions varies significantly. The average NPEs were lower in the middle and near edges. This is because the majority of photons received by the SiPMs come from the total reflection inside the scintillator. The number of photons reflected from the front and rear sides of the scintillator (along Z direction in Fig.~\ref{fig:2}) is relatively consistent with the current geometry and depends smoothly on the azimuth of each SiPM. The photons reflected from the upper and lower sides (along Y-direction in Fig.~\ref{fig:2}) will also hit the SiPMs after one total internal reflection, which means that the closer the SiPM is to the edge, the more photons are received. In addition, the corners of the light guide were fixed to the cassette. Therefore, a light-absorbing layer was added at these positions in the simulation, resulting in a decrease in photon collection efficiency by the SiPMs at both ends.
\begin{table}
\begin{center}
\caption{Average photoelectron of 16 SiPM arrays on one side}
\begin{tabular}{|c|c|c|c|c|}
 \hline 
 
Beam &500 MeV/u  $^{238}$U  & 600 MeV/u  $^{40}$Ar & 310 MeV/u  $^{40}$Ar & 600 MeV/u  $^{12}$C  \\
\hline 
Scintillator thick/mm& 0.8 & 0.8 & 0.8& 5 \\
 \hline  
Energy loss/MeV& 1814 & 64.29 & 89.65& 43.98 \\
 \hline  
Quenching factor & 0.203 & 0.284 & 0.262& 0.615 \\
 \hline 
Average photoelectron &  3103.5 & 154.9 & 198.9 & 171.7 \\
\hline  
\end{tabular}
\label{fig:t1}
\end{center}
\end{table}

\subsection{Waveform simulation}
\begin{figure}[htbp]
\begin{minipage}[t]{0.5\linewidth}
\centering
\includegraphics*[width=1\textwidth]{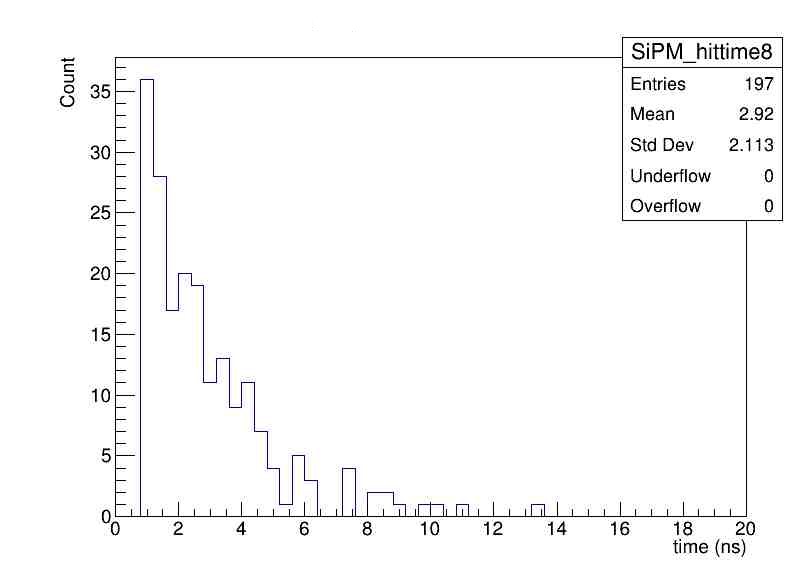}
\centerline{(a)}
\end{minipage}
\hfill
\begin{minipage}[t]{0.5\linewidth}
\centering
\includegraphics*[width=1\textwidth]{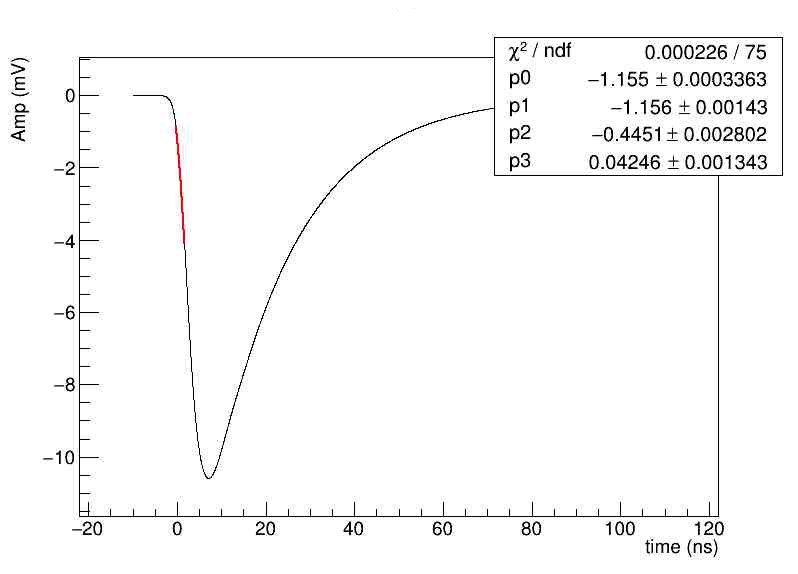}
\centerline{(b)}
\end{minipage}
\caption{(a) Arrival time of photoelectrons received by a SiPM in an event; (b) Output signal}
\label{fig:5}
\end{figure}

The single photon–electron (SPE) response waveform $v_{i (t)}$ can be simulated using an experimentally measured SPE signal of the SiPM combined with polynomial fitting. The output signal of the SiPM is the superposition of all the SPE signals in one event, which is given in Formula ~\ref{equation:BTV}:

\begin{equation}
V(t)=\sum_{i=1}^{NPE}v_i\{t-t_{transit}-(t_0)_i\}
\label{equation:V(t)}
\end{equation}
where NPE represents the total number of photoelectrons received by a SiPM, $t_{transit}$ is the single photon transit time, and $t_0$ represents the time that the single photon arrives at the SiPM. The value of $t_{transit}$ is sampled from a Gaussian function with a mean value of 0 (when analyzing time resolution, the specific value of $t_{transit}$ does not affect the analysis result) and sigma value of 1037 ps (which is taken from the experimentally measured single-photon time resolution of SiPM). Fig.~\ref{fig:5} shows the photon arrival time distribution and the simulated output waveform of SiPM 8 when 310 MeV/u 40Ar beam passes through the center of the T0 detector.

\subsection{Time performance}
\begin{figure}[htbp]
\begin{minipage}[t]{0.5\linewidth}
\centering
\includegraphics*[width=0.98\textwidth]{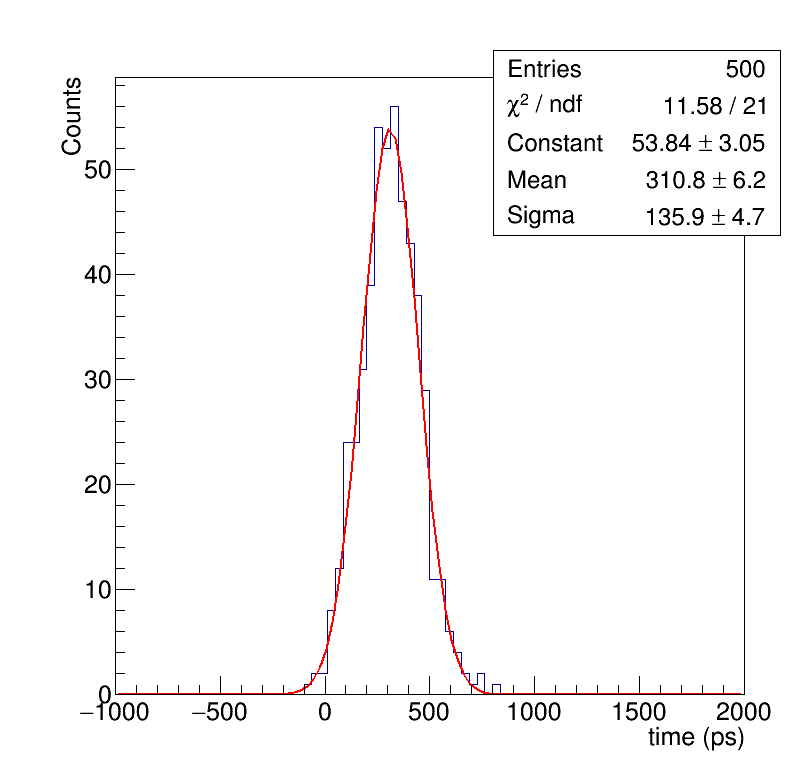}
\centerline{(a)}
\end{minipage}
\hfill
\begin{minipage}[t]{0.5\linewidth}
\centering
\includegraphics*[width=1\textwidth]{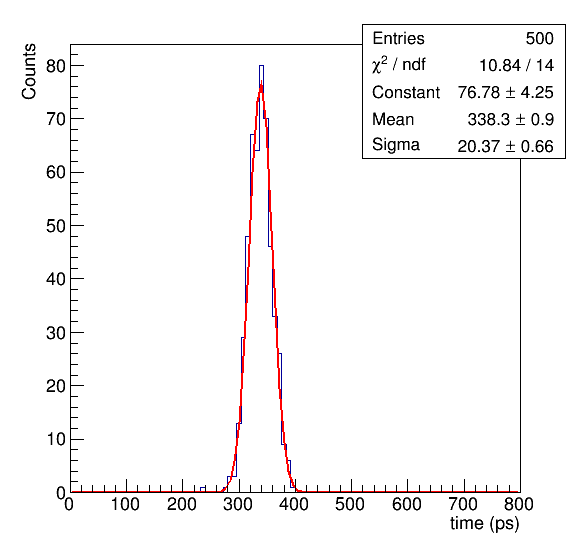}
\centerline{(b)}
\end{minipage}
\caption{Simulated single-channel time resolution and system time resolution of T0 detector}
\label{fig:6}
\end{figure}

The timing information $t_i$ for each SiPM is obtained by taking a constant fraction discrimination (CFD) of the leading edge of the simulated SiPM output signal at 20\% of the maximum amplitude. Then, the reference time of the T0 detector can be defined as the arithmetic mean of the times of all 32 SiPMs:
\begin{equation}
T_{ref}=(\sum_{i=1}^{32}t_i)/32
\label{equation:T0Res}
\end{equation}
where we do not consider the time difference between photon transmission, SiPM response, and electronic readout. This time difference is constant and does not affect the time resolution of T0 detector. Fig.~\ref{fig:6} shows the simulated distributions of the output time when the 310 MeV/u $^{40}$Ar beam passed through the center of the T0 detector. Fig.~\ref{fig:6} (a) is the time distribution of one of the 32 SiPMs, and Fig.~\ref{fig:6} (b) is the simulated distribution of $T_{ref}$ as defined by Equation~\ref{equation:T0Res}. Their time jitters obtained by Gaussian fitting were 135.9 ps and 20.37 ps, respectively.

$T_{ref}$ defined by Equation~\ref{equation:T0Res} provides the best estimation of the reference time of beam crossing if all 32 channels of SiPMs show the same timing resolution. However, according to the simulation results shown in Fig.~\ref{fig:4}, the number of photoelectrons received by different SiPMs vary significantly. Therefore, the time resolution of each channel is different, and the arithmetic mean timing of 32 SiPMs will not be the best $T_{ref}$ estimation. For this reason, the weighted mean of timing is introduced to determine $T_{ref}$. The weighting factor for each SiPM was $w_1=1/\sigma_1^2$, $w_2=1/\sigma_2^2$, ..., $w_{32}=1/\sigma_{32}^2$  respectively, where $\sigma_1$, $\sigma_2$, ..., $\sigma_{32}$ represent the timing jitter of each SiPM. The timing of T0 system using the weighted mean method can be calculated as
\begin{equation}
T_{refcor2}= (\sum_{i=1}^{32}w_it_i)/(\sum_{i=1}^{32}w_i)
\label{equation:T0Res2}
\end{equation}

\begin{figure}[htbp]
\centering
\includegraphics[width=1.\textwidth]{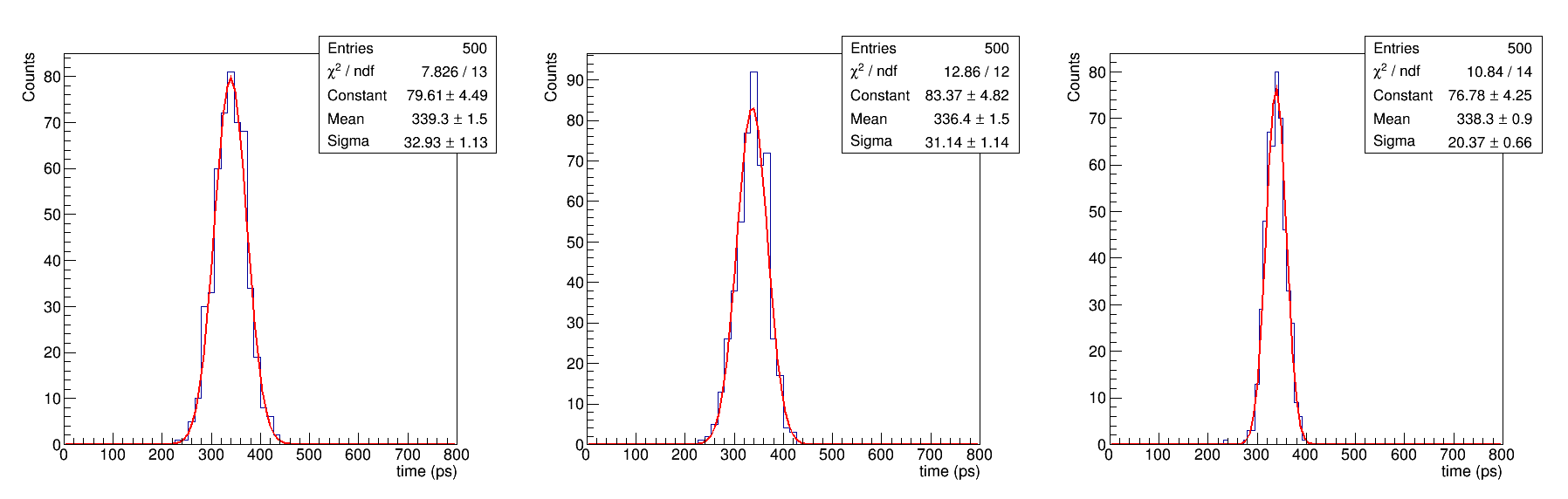}
\centerline{(a)\qquad\qquad\qquad\qquad\qquad(b)\qquad\qquad\qquad\qquad\qquad(c)}
\caption{The reference time of T0 using the weighted mean method}
\label{fig:7}
\end{figure}
Fig.~\ref{fig:7} shows the time distribution of the T0 system using the weighted mean method, where graph (a) and (b) represent the average time of 16 SiPMs on each side, respectively, while graph (c) represents $T_{refcor2}$ according to Equation ~\ref{equation:T0Res2}. When the beam crosses the center of T0, the weighted mean method does not show significant improvement in timing compared to the simple average method (Equation~\ref{equation:T0Res}). The reason is that although the NPEs received by different SiPMs are different, they are large enough so that, the timing of SiPM does not change much. Therefore, Equations~\ref{equation:T0Res} and~\ref{equation:T0Res2} give similar results. However, if the beam crossing point is far away from the center of the T0 detector, the results of different algorithms will show distinct differences, which will be described in later chapters.

\section{Radiation resistance}
When the heavy-ion beam passes through the plastic scintillator, a large amount of energy is deposited, which inevitably causes radiation damage to the plastic scintillator. Moreover, its performance decreases after long-term use. Therefore, it is necessary to study the irradiation performance of the T0 detector and evaluate its service life. Chong et al.~\cite{wucho} studied the irradiation performance of an EJ-200 plastic scintillator. It is shown that when the radiation dose exceeded $1.44\times10^{4}$ Gy, the transmission spectrum deteriorated, and the light output decreased rapidly. The maximum service life of the scintillator can be calculated using $1.44\times10^{4}$ Gy as the upper limit of the radiation dose. According to Geant4 simulation, when 600 MeV/u $^{12}$C beam and 500 MeV/u $^{238}$U beam passed through the EJ-200 plastic scintillators with thicknesses of 5 mm and 0.8 mm, the average energy depositions were 43.98 MeV and 1814 MeV, respectively. The corresponding radiation doses of $^{12}$C and $^{238}$U were 0.0175 Gy/s and 4.523 Gy/s, respectively, with a beam intensity of $10^{6}$ per second and a beam spot diameter of 5 mm. According to the beam extraction design at CEE, the beam bunch lasts for 2 s, and the average interval time between bunches was $\sim$30 s. Therefore, the service lives of the scintillator under $^{12}$C and $^{238}$U beam irradiation were approximately 6857.1 h and 26.5 h, respectively.
\begin{figure}[htbp]
\centering
\includegraphics[width=1.\textwidth]{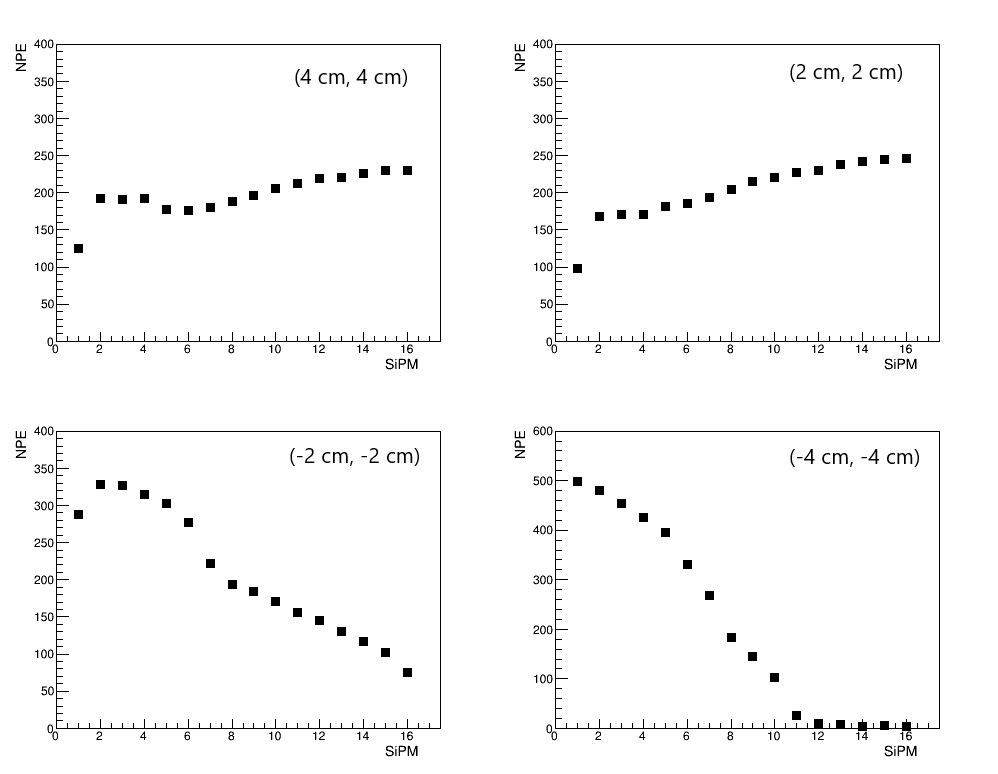}
\caption{The number distribution of photoelectrons received by 16 SiPMs on the right side of T0 when the 310 MeV/u $^{40}$Ar beam passes through different positions}
\label{fig:9}
\end{figure}

A two-dimensional moving platform was introduced to improve the service life of T0, particularly for heavy-nucleus beams. The beam-hit position on the scintillator can be adjusted during the experiment to avoid excessive irradiation. However, when the beam incident position on the T0 detector changes, the number of photoelectrons received by the SiPM arrays on both sides also changes. Therefore, it is necessary to study the timing performance of T0 as a function of the various beam-hit positions.\par

A two-dimensional Cartesian coordinate system was established using the center of T0 as the coordinate origin, and the X-and Y-axes as the horizontal and vertical axes, respectively, as shown in Fig~\ref{fig:2}. We selected four points on the scintillator: (-4 cm, -4 cm), (-2 cm, -2 cm), (2 cm, 2 cm), and (4 cm, 4 cm). The number of photons received by the SiPMs was simulated, and the time resolution of T0 was evaluated. Fig.~\ref{fig:9} shows the distribution of the number of photoelectrons received by the 16 SiPMs on the left side (+X direction) of T0 when the 310 MeV/u $^{40}$Ar beam passed through the four positions. It can be seen that the number of photoelectrons received by each SiPM changes significantly when the beam is incident off-center compared to when it is incident centrally. The results from the hit position (-4 cm, -4 cm) show that the number of photoelectrons changes rapidly and is close to 0 from the 11th SiPM. This is because starting from the $11^{th}$ to the $16^{th}$ SiPM, most of the photons reaching the light-guide-air-SiPM interface will be totally reflected and will not reach the SiPM.

\begin{figure}[htbp]
\centering
\includegraphics[width=1.\textwidth]{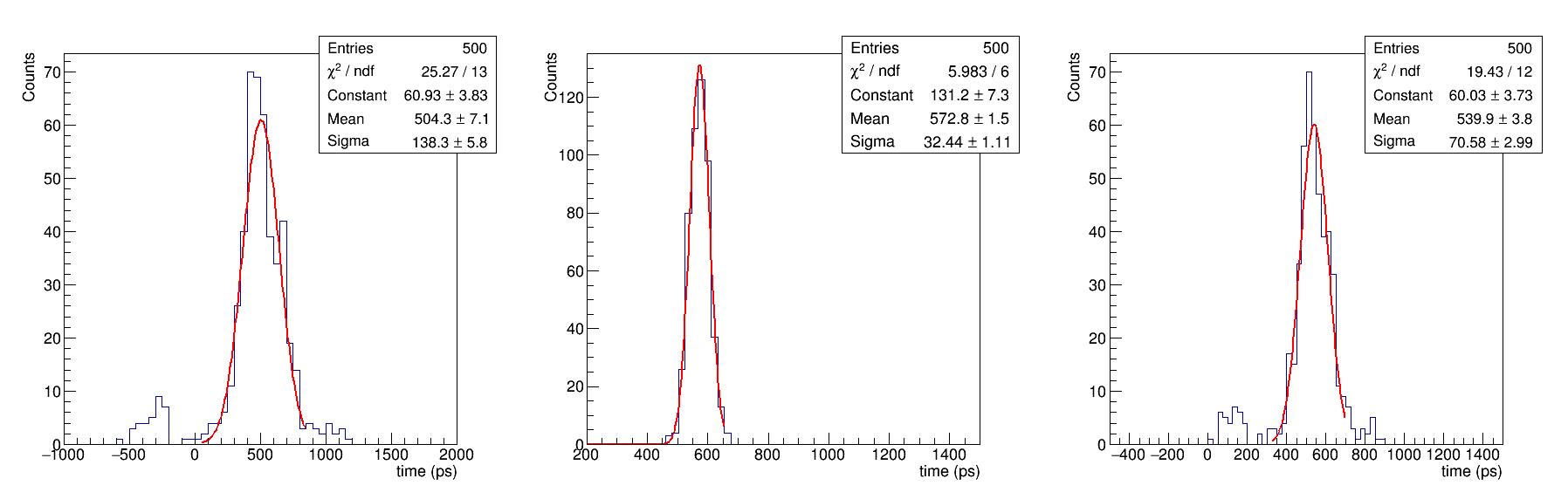}
\centerline{(a)\qquad\qquad\qquad\qquad\qquad(b)\qquad\qquad\qquad\qquad\qquad(c)}
\caption{The reference time of T0 using the arithmetic average algorithm}
\label{fig:10}
\end{figure}

\begin{figure}[htbp]
\centering
\includegraphics[width=1.\textwidth]{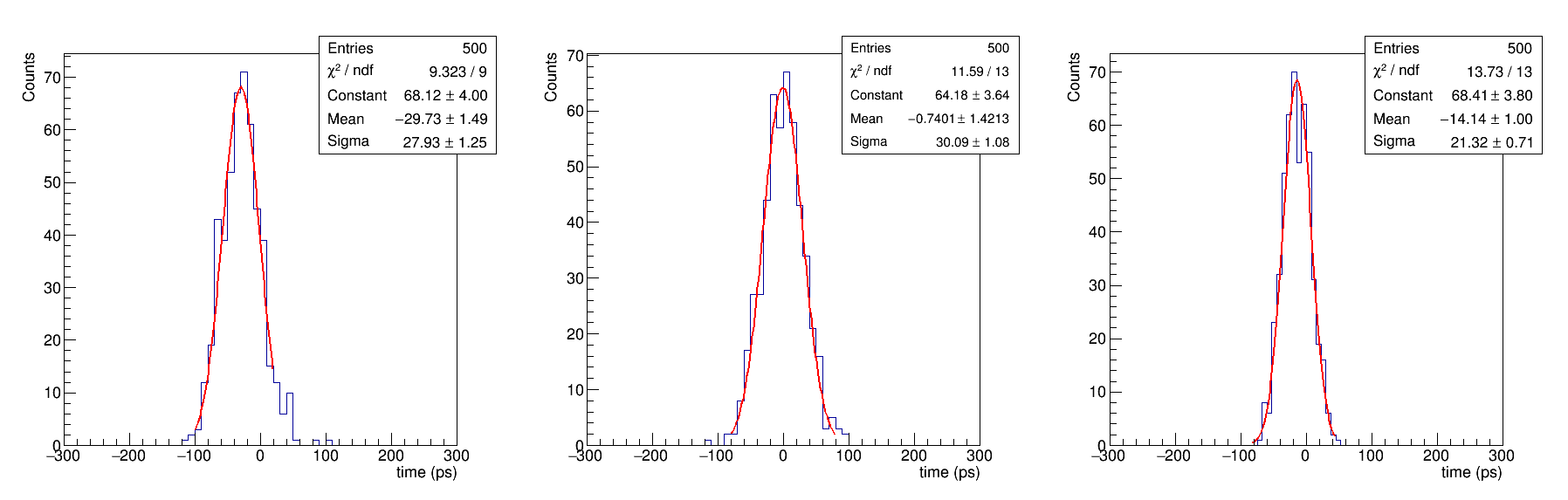}
\centerline{(a)\qquad\qquad\qquad\qquad\qquad(b)\qquad\qquad\qquad\qquad\qquad(c)}
\caption{The reference time of T0 using the weighted mean method}
\label{fig:11}
\end{figure}

The hit position (-4 cm, -4 cm) shows the most significant difference in the photoelectron number distribution. Therefore, the time performance of the T0 detector was studied when the beam hit the scintillator at this position. The simulation results are shown in Figs.~\ref{fig:10} and Fig~\ref{fig:11}, where the reference time of T0 is calculated using the arithmetic average method (Equation~\ref{equation:T0Res}) and the weighted mean method (Equation~\ref{equation:T0Res2}), respectively. Figs.~\ref{fig:10} (a) and Fig~\ref{fig:11} (a) illustrate the average time distribution of 16 SiPMs on the left side of the T0 detector, while Figs.~\ref{fig:10} (b) and Fig~\ref{fig:11} (b) show the results of 16 SiPMs on the right side. The mean time distributions of all 32 SiPMs are shown in Figs.~\ref{fig:10} (c) and Fig~\ref{fig:11} (c). The time resolutions in Fig.~\ref{fig:10} (a), (b) and (c) were 138.3, 32.44, and 70.58 ps, respectively, using the arithmetic average method. When the beam passes away from the center of T0, the time performance of T0 is degraded compared to the simulation result in Fig.~\ref{fig:6}. However, using the weighted mean method, the timing performance of T0 remains similar to that of the central beam incidence, as depicted in Fig.~\ref{fig:11}. This result verifies the feasibility of extending T0 lifetime by changing the beam hit position on the plastic scintillator. It is worth noting that even within the radiation dose limit, the plastic scintillator exhibits non-negligible fluorescence yield loss and attenuation in light propagation. Therefore, it is necessary to arrange the beam incident point and order reasonably to minimize the impact of the irradiated area as much as possible.

\section{Readout electronics}
\begin{figure}[htbp]
\centering
\includegraphics[width=.6\textwidth]{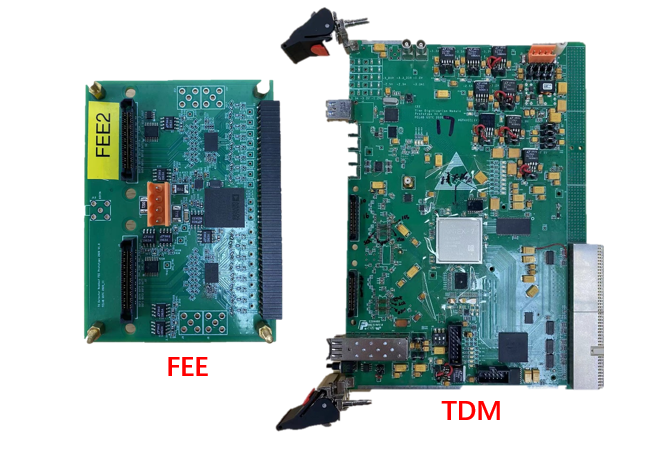}
\caption{FEE and TDM}
\label{fig:12}
\end{figure}

T0 readout electronics employ a front-end electronic (FEE) module utilising NINO ASIC\cite{anghinolfi2003nino,anghinolfi2004nino} and a time-digitization-module (TDM) with FPGA TDC to achieve multi-channel and high-time-accuracy measurements, as reported in~\cite{cao2022prototype}. Fig.~\ref{fig:12} shows a photograph of the FEE and TDM prototypes. Each FEE integrates two NINO chips, which can amplify and discriminate 16 SiPM signals. The time-over-threshold (TOT) output by the FEE was transmitted to the TDM through coaxial cables to complete the time measurement. In addition to the time measurement circuit, a temperature compensation system was implemented in the FEE and TDM modules. The temperature was measured using a temperature sensor on the SiPM board, and the voltage was provided by a power circuit on the FEE. Temperature compensation logic is implemented in the FPGA on the TDM module to control the voltage supply to the SiPMs

\begin{figure}[htbp]
\begin{minipage}[t]{0.5\linewidth}
\centering
\includegraphics*[width=1\textwidth]{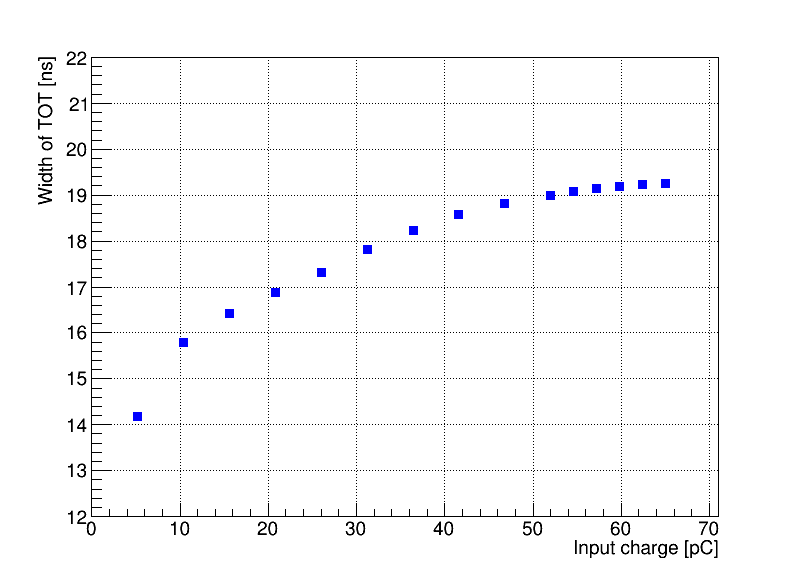}
\caption{Width of TOT output by NINO varies with input signal charge}
\label{fig:13}
\end{minipage}
\hfill
\begin{minipage}[t]{0.5\linewidth}
\centering
\includegraphics*[width=1.\textwidth]{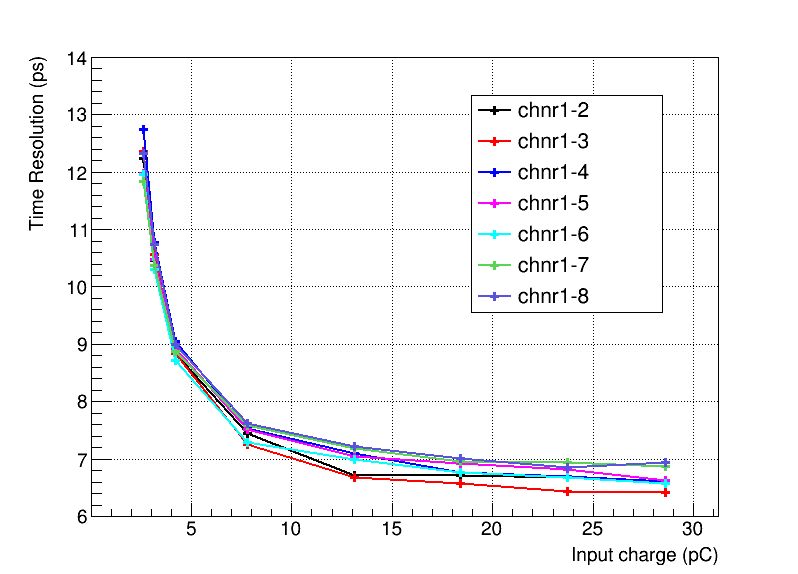}
\caption{Time jitter in different channels}
\label{fig:14}
\end{minipage}
\end{figure}

The NINO ASIC is a dedicated chip for multi-gap resistive plate chamber (MRPC) operation, and its dynamic range is 200 fC–2 pC. The Hamamatsu S13360-3025 series SiPM has a higher charge than the MRPC signal. For example, at a working voltage of 62 V and an NPE of 150, the output charge of the SiPM was approximately 15 pC. A 10 nF capacitor is connected in series with the NINO to filter and attenuate the signal in order to reduce the amount of input charge. The readout electronics were tested using an 81180A signal source to generate a SiPM signal with a rising edge of 2 ns and a falling edge of 50 ns. Fig.~\ref{fig:13} shows the width of the TOT output measured by NINO under different input charges. In the test range, the input and output of the NINO exhibited a good linear relationship. However, when the input charge is greater than 30 pC, the crosstalk between the adjacent channels of NINO becomes significant.

The time jitter of different channels was obtained using the same test setup, and the results are shown in Fig.~\ref{fig:14}. In the dynamic range of 3–30 pC, the time jitter of the different channels was less than 15 ps.

\section{Performance test}
\begin{figure}[htbp]
\begin{minipage}[t]{1.\linewidth}
\centering
\includegraphics*[width=.8\textwidth]{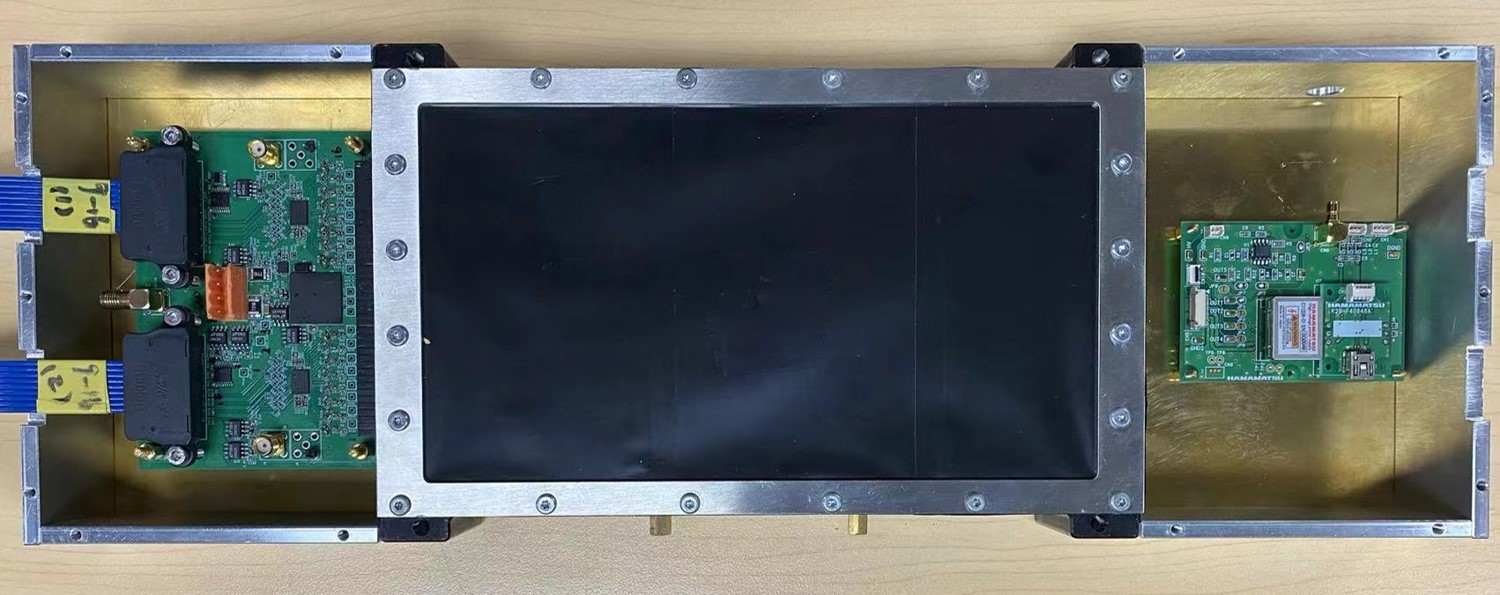}
\centerline{(a)}
\end{minipage}
\hfill
\begin{minipage}[t]{1.\linewidth}
\centering
\includegraphics*[width=.4\textwidth]{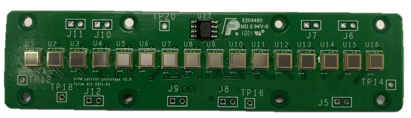}
\centerline{(b)}
\end{minipage}
\caption{T0 prototype}
\label{fig:15}
\end{figure}

The T0 prototype was developed and tested in the laboratory. Fig.~\ref{fig:15} (a) shows a photograph of the prototype. The plastic scintillator was placed in a light-tight cassette, and the SiPMs were welded to the SiPM board, which was also sealed in the cassette. Fig.~\ref{fig:15} (b) shows an SiPM board with 16 assembled SiPMs, and the temperature sensor is shown in black on the PCB. The outer frame of the cassette was made of aluminum, and the front and back faces were masked by black Kapton films of 25 $\mu m$ thick. Two types of readout electronics were installed on the left and right sides of the T0 prototype (as shown in Fig.~\ref{fig:15} (a)) to perform different tests. The left end was equipped with NINO FEE electronics, which can realize the timing information measurement of 16-channel SiPM signals. On the right side was a set of Hamamatsu C12332-01 electronics, which is used to realize signal charge measurement of a single-channel SiPM.

\subsection{SiPM test}
\begin{figure}[htbp]
\centering
\includegraphics[width=.5\textwidth]{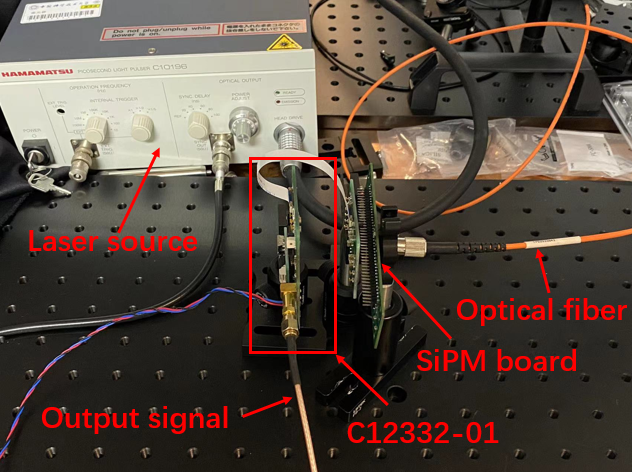}
\caption{SiPM test system}
\label{fig:16}
\end{figure}

The gain of the SiPM at different bias voltages was tested while the temperature was kept at 25.5$^{\circ}$C. The bias voltage was provided by a C12332-01 module. The Hamamatsu C10196 laser source was used to illuminate the SiPM under testing using an optical fiber. The SiPM output signal was sent to the C12332-01 module for amplification and processing. Fig.~\ref{fig:16} shows a photograph of the test setup.

\begin{figure}[htbp]
\begin{minipage}[t]{.5\linewidth}
\centering
\includegraphics*[width=1.\textwidth]{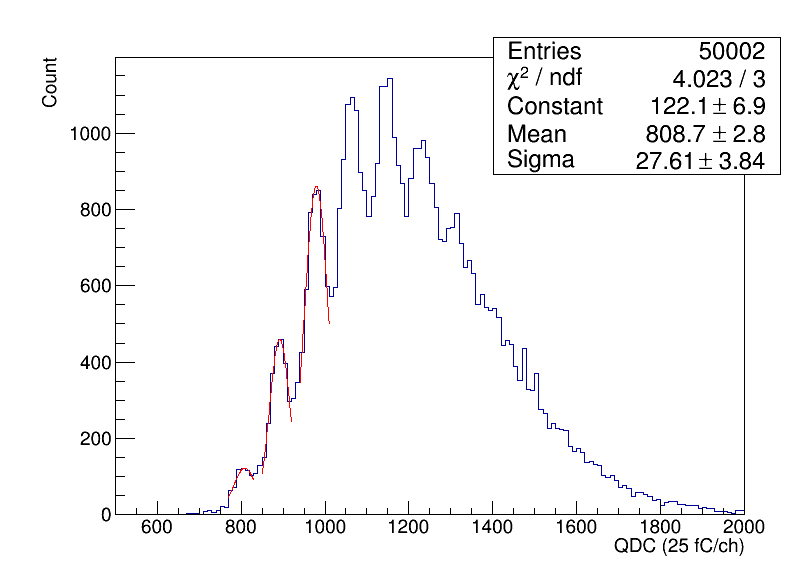}
\centerline{(a)}
\end{minipage}
\hfill
\begin{minipage}[t]{.5\linewidth}
\centering
\includegraphics*[width=1.\textwidth]{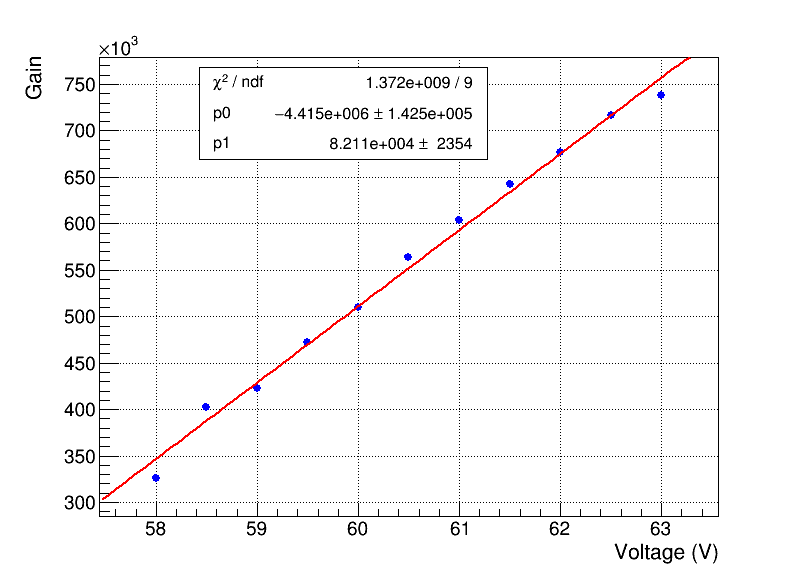}
\centerline{(b)}
\end{minipage}
\caption{SiPM gain via bias voltage}
\label{fig:17}
\end{figure}

The charge of the SiPM signal was measured using a charge-to-digital converter (QDC) (CAEN
V965A, 25 fC/ch). Fig.~\ref{fig:17} (a) shows the charge spectrum when the bias voltage was set to 62 V, with the first peak representing the pedestal and the other peak representing single and multiple photoelectrons.
\begin{equation}
Gain = \frac{(mean_{2}-mean_{1})\times 0.025}{1.6\times10^{-7}\times 20 }
\label{equation:gain}
\end{equation}
The gain of one photoelectron can be calculated by Equation~\ref{equation:gain}, where the $mean_1$ and $mean_2$ represent the peak value of single and dual photoelectron charge spectrum respectively, and the value of 20 represents the amplification factor of C12332-01 module. Fig.~\ref{fig:17} (b) depicts the dependence of SiPM gain on the bias voltage, where a linear correlation is obvious.
\subsection{Heavy ion beam test}
\begin{figure}[htbp]
\centering
\includegraphics[width=.8\textwidth]{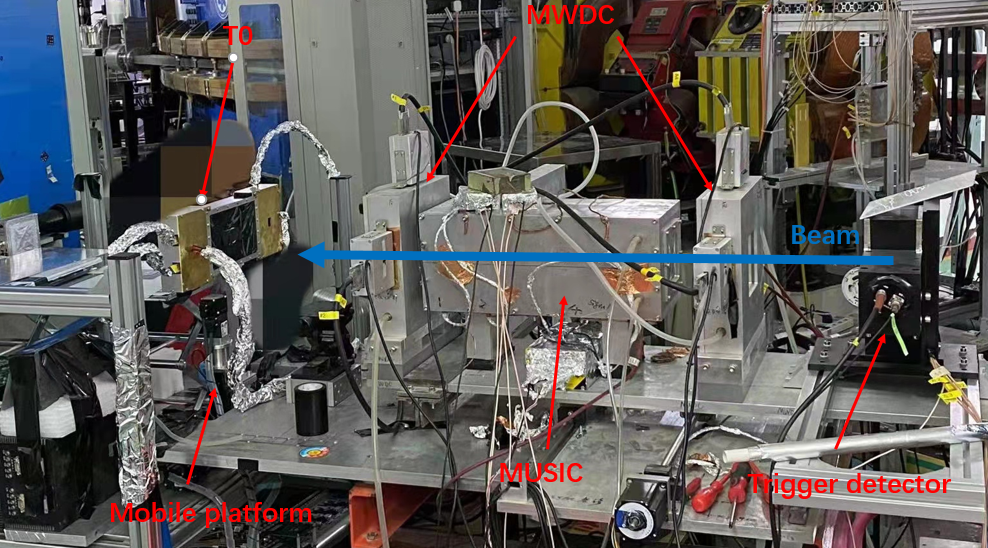}
\caption{Heavy ion beam test detector layout}
\label{fig:18}
\end{figure}

We performed a heavy-ion beam test at the external-target facility (ETF) of the second radioactive ion beam line in Lanzhou (RIBLL2) at HIRFL-CSR to verify whether the T0 prototype meets the performance requirements of the CEE. $^{40}$Ar beam with a kinetic energy of 320 MeV/u was used here. The scintillator thickness of the T0 prototype under test was 0.8 mm. Fig.~\ref{fig:18} depicts the layout of the test setup at RIBLL2-ETF. Heavy-ion beams from the left side successively passed through a trigger detector, two multi-wire drift chamber (MWDC) trackers, and a multi-sampling ionization chamber (MUSIC) and then reached the T0 prototype. The $^{40}$Ar beam would lose its energy when crossing these detectors and hit the T0 prototype with an energy of $\sim$ 310 MeV/u. The T0 prototype was installed on a 2-D moving platform, allowing X-Y movement in the range of (-4 cm, -4 cm) to (4 cm, 4 cm)

The readout electronics used in the heavy-ion beam test are similar to those used in the laboratory, as shown in Fig~\ref{fig:15}. The difference is that the NINO electronics employed a self-trigger mode for readout triggering and data selection in the beam test. A specific trigger logic that required all 16 SiPMs on one side of the T0 prototype to fire simultaneously (within a short time window of XX ns) was implemented to trigger a valid heavy-ion event. A copy of this trigger was sent to the QDC for the SiPM signal charge measurement.
\begin{figure}[htbp]
\begin{minipage}[t]{.5\linewidth}
\centering
\includegraphics*[width=1.\textwidth]{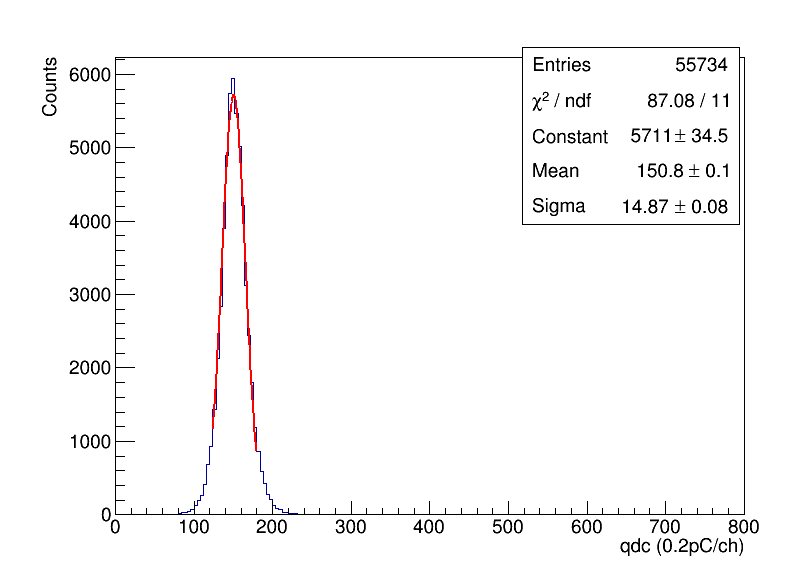}
\centerline{(a) pedstal}
\end{minipage}
\hfill
\begin{minipage}[t]{.5\linewidth}
\centering
\includegraphics*[width=1.\textwidth]{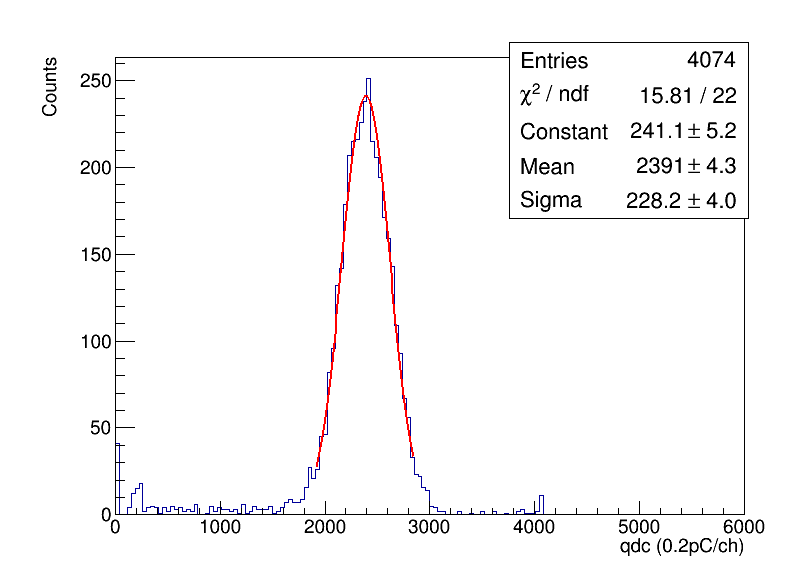}
\centerline{(b) beam signal}
\end{minipage}
\caption{Charge spectrum measured by QDC}
\label{fig:19}
\end{figure}

\begin{figure}[htbp]
\begin{minipage}[t]{.5\linewidth}
\centering
\includegraphics*[width=1.\textwidth]{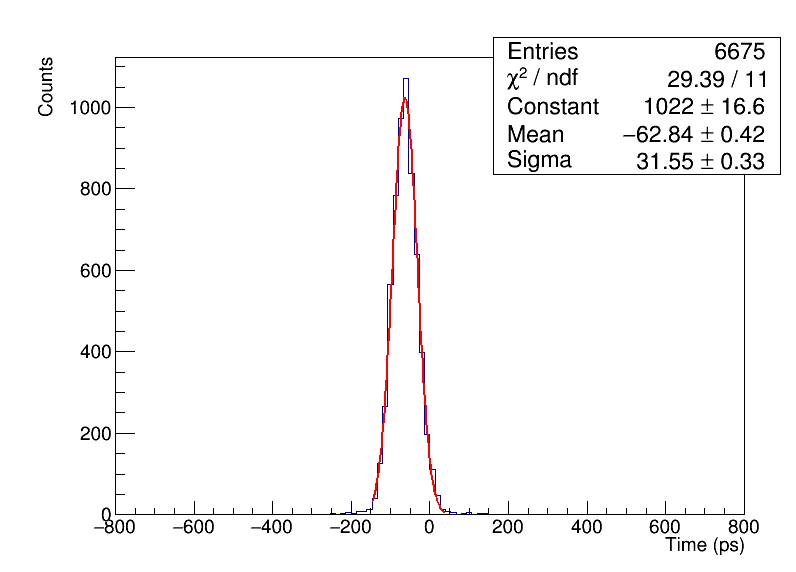}
\centerline{(a)}
\end{minipage}
\hfill
\begin{minipage}[t]{.5\linewidth}
\centering
\includegraphics*[width=1.\textwidth]{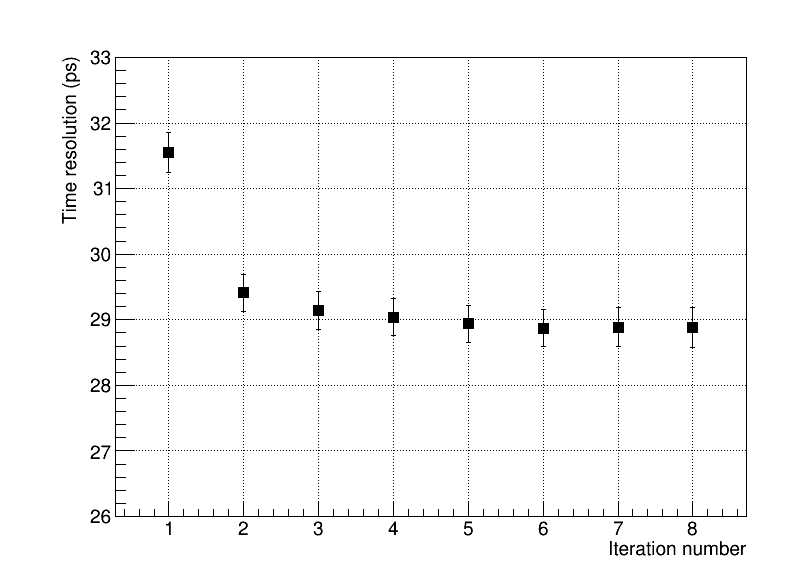}
\centerline{(b)}
\end{minipage}
\caption{T0 system time resolution calibrated by 16 SiPMs. (a) Time resolution after T-TOT correction. (b) Time resolution via a number of iterations.}
\label{fig:20}
\end{figure}
The SiPMs worked at a bias voltage of 62 V. Fig.~\ref{fig:19} shows the charge spectrum of SiPM 8, measured by QDC (CAEN V965A,0.2 pC/ch), where plots (a) and (b) represent the pedestal and signal charges with and without beam crossing, respectively. It can be calculated that with a $^{40}$Ar beam crossing, the average number of photoelectrons received by SiPM 8 was:
\begin{equation}
NPE = \frac{(2390-150.7)\times 0.2}{1.6\times10^{-7}\times 20\times 6.77\times10^5 }=206.7
\end{equation}
where $6.77\times10^5$ represents the gain of SiPM 8 at a bias voltage of 62 V, according to the gain test results in Fig.~\ref{fig:17}.

The reference time of T0 can be self-calibrated using the time information of the 16 SiPMs measured by TDM:
\begin{equation}
T_{ref} = \frac{\sum_{i=1}^{16}(t_i-tcor_i-t0)/\sigma_i^2}{\sum_{i=1}^{16}1/\sigma_i^2 }
\label{equation:T0}
\end{equation}
where $tcor_i$ is the T-TOT correction term, $\sigma_i$ represents the time resolution of $i_{th}$ SiPM and t0 is the weighted mean time of 16 SiPMs, defined as:
\begin{equation}
t0=\frac{\sum_{i=1}^{16}(t_i-t_{cori})/\sigma_i^2}{\sum_{i=1}^{16}1/\sigma_i^2 }
\label{equation:T02}
\end{equation}
In Equation~\ref{equation:T0} and Equation~\ref{equation:T02},  $\sigma_i$ and t0 are interrelated parameters, and this correlation can be eliminated by using recursive calibrations. Fig.~\ref{fig:20} (a) shows the timing distribution of T0 after the first round of calibration, and a time resolution of 31.6 ps was obtained. Fig.~\ref{fig:20} (b) shows the time resolution of the T0 as a function of iteration times. It can be seen that as the number of iterations increases, the time resolution of T0 gradually approaches ∼28.9 ps.

As described in Section 3.4, we installed a two-dimensional moving platform so that the beam incident position on the T0 detector could be adjusted to extend the service life of T0. This feature was tested using heavy ion beams. Table~\ref{fig:t2}  lists the number of photoelectrons received by SiPM 8 under different beam incident positions (notice that SiPM 8 was located on  the positive X side). Different columns and rows in the table denote different hit positions along the X and Y directions, respectively. The time resolutions of T0 at different beam incident positions after 10 calibration iterations are presented in Table~\ref{fig:t3}. It should be noted that this time resolution was achieved by using 16 SiPMs. Thus, with all 32 SiPMs functional, the T0 detector can provide an even better timing performance.
\begin{table}
\begin{center}
\caption{The number of photoelectrons received by SiPM 8 under different beam incident positions}
\begin{tabular}{|c|c|c|c|c|c|c|}
 \hline 
Position (X/Y) (mm)  &10 & 0 &-10 & -20 &-30 &-40  \\
\hline 
0  & 216.7 &  206.7 & 189.8 & 188.7 & 182.1&177.2 \\
 \hline  
-10&       & 189.6&       &  189.3 &      &  \\
 \hline 
-20& 190   & 184.8&       & 185.5  &      &174.6 \\
 \hline  
-30 & 187.8 & 183.6 &     & 178.6  &      &173.4 \\
 \hline 
\end{tabular}
\label{fig:t2}
\end{center}
\end{table}

\begin{table}
\begin{center}
\caption{Time resolution of T0 at different beam incident positions (unit:ps)}
\begin{tabular}{|c|c|c|c|c|c|c|}
 \hline 
Position (X/Y) (mm) &10 & 0 &-10 & -20 &-30 &-40  \\
\hline  
0  & 29.63 &  28.9& 27.8 & 26.29 & 24.79&24.08 \\
 \hline 
-10&       & 29.80&       &  29.71 &      &  \\
 \hline  
-20& 29.96  & 28.80&       & 28.42  &      &24.45 \\
 \hline  
-30 & 27.88 & 26.44 &     & 24.17  &      &23.46 \\
 \hline 
\end{tabular}
\label{fig:t3}
\end{center}
\end{table}

\section{Summary}
The TOF system plays an important role in identifying charged final-state particles in the CEE. The T0 detector, which provides a precise reference time for the TOF system, is located upstream of the target along the beam line. We proposed a T0 detector based on a thin plastic scintillator coupled with 32 SiPMs. The readout electronics of T0 employed a FEE utilising a NINO ASIC and a TDM with FPGA TDC. This study detailed the design of T0 and investigated its performance through GEANT4 simulations, including energy deposition in plastic scintillators, photoelectrons received by SiPMs, and reference time determination. The radiation resistance of T0 was also studied via simulation, and a 2-D moving platform was added to improve the service life of T0 by changing the beam incident position. T0 exhibited excellent time performance using the weighted mean method under different beam incident positions. Finally, a T0 prototype with 16 SiPMs was tested using a $^{40}$Ar beam at RIBLL2-ETF. The test results demonstrated that with our design, the T0 prototype achieved a time resolution better than 30 ps with only 16 SiPMs, which was consistent with the GEANT4 simulation results and fulfilled the CEE requirement.

\section*{Acknowledgments}
The authors thank the high energy physics group of USTC. This project is supported by the National Natural Science Foundation of China under grant No. U11927901,12205296 and 11975228, the State Key Laboratory of Particle Detection and Electronics under grant No. SKLPDE-ZZ-202202 and the USTC Research Funds of the Double First-Class Initiative under Grant No.WK2030000052.  Special thank goes to Professor Qing Luo from National Synchrotron Radiation Laboratory (China) for help with FEM calculation (ANSYS).

\bibliography{mybibfile}
\end{CJK}
\end{document}